\def\lsim{\mathrel{\rlap{\lower4pt\hbox{\hskip1pt$\sim$}}
    \raise1pt\hbox{$<$}}}         
\def\gsim{\mathrel{\rlap{\lower4pt\hbox{\hskip1pt$\sim$}}
    \raise1pt\hbox{$>$}}}         

\def\ie{\hbox{\it i.e. }}

\def\eg{\hbox{\it e.g. }}

\def\etal{\hbox{\it et al. }}

\def\g2{ GeV$^2$}
\def\xi2{$\chi^2_{d.o.f}$}
\def\underarrow#1{\mathrel{\mathop{\longrightarrow}\limits_{#1}}}

\documentstyle[12pt]{article}   
\input epsf
\begin{document}
\begin {titlepage}

February 1998 \hfill LYCEN 9808 \\
\vskip 0.8cm
\centerline {\bf INTERPOLATING BETWEEN SOFT AND HARD DYNAMICS}
\centerline {\bf IN DEEP INELASTIC SCATTERING}
\vskip 2cm
\centerline{\bf P. Desgrolard ({\footnote{E-mail: desgrolard@ipnl.in2p3.fr}}),
L. Jenkovszky ({ \footnote{E-mail: jenk@gluk.apc.org}}), 
F. Paccanoni ({ \footnote{E-mail: paccanoni@pd.infn.it}   })} 
\vskip 1cm
($^1$) {\it Institut de Physique Nucl\'eaire de Lyon, IN2P3-CNRS et
Universit\'e Claude Bernard, F69622 Villeurbanne Cedex, France.}

($^2$) {\it Bogolyubov Institute for Theoretical Physics, Kiev-143,
Ukraine.}

($^3$) {\it Dipartimento di Fisica, Universit\'a di Padova,
Istituto Nationale di Fisica Nucleare, Sezione di Padova, 
via F.Marzolo, I-35131 Padova, Italy.}

\vskip 3cm
\centerline {\bf Abstract}
\vskip 1cm

An explicit model for the proton structure function is
suggested, interpolating between low-$Q^2$ vector meson dominance and Regge
behavior,                                                            
on the one hand, and the high-$Q^2$ solution of the 
Gribov-Lipatov-Altarelli-Parisi  evolution
equation, on the other hand. The model is fitted to the
experimental data in a wide range of the kinematical variables with
emphasis on the low-$x$ HERA data. The boundaries, transition region 
and interface between various regimes are quantified.

\end{titlepage}

  {\bf 1 INTRODUCTION }

\medskip

In deep inelastic scattering the dynamics of low - and high
virtualities, $Q^2$ is usually treated in a disconnected way, by
using different methods. The structure function (SF) $F_2(x,Q^2)$ at
small $Q^2$ (and small $x$, where $x$ is the fraction of the momentum 
carried by a parton) is known to be Regge-behaved and satisfying 
vector meson dominance (VMD) with the limit 
$F_2(x,Q^2)\underarrow{Q^2\rightarrow 0}0$, imposed by gauge
invariance. At large $Q^2$, on the other hand, $F_2(x,Q^2)$ obeys the
solutions of the
Gribov-Lipatov-Altarelli-Parisi (GLAP) evolution equation [1].

One important problem remains open: where do these two regimes meet
and how do they interpolate? In the present paper we seek answers to
these questions.

For definiteness, we deal with the proton SF to be denoted $F_2$.
Our emphasis is on the small $x$ region, dominated by
gluodynamics.  The valence quark contribution will be added at
large-$x$ in a phenomenological way to make the fits complete.

The forthcoming presentation has also an important aspect
relevant to quantum chromodynamics (QCD), namely in clarifying
the range of applicability and the interface between the GLAP and the
Balitsky-Fadin-Kuraev-Lipatov (BFKL) [2] evolutions.  While the GLAP
equation describes the evolution of the SF in $Q^2$ starting from a
given $x-$ dependence, the BFKL evolution means variation
of the SF in $x$ for fixed $Q^2$,  both implying large enough $Q^2$
for the perturbative expansion to be valid. QCD leaves flexible the
relevant limits and boundaries. Moreover, the onset of their
asymptotic solutions depends on details of the calculations.
In this paper we try to make some of these limits explicit and
quantitative.

HERA is an ideal tool to verify the above theories. The relevant data
extend over a wide range of $Q^2$ - a fruitful test field for the 
GLAP evolution, on one hand, and to low enough $x$, where the SF is 
dominated by a Pomeron contribution, expected to be described by the
BFKL evolution (see below).

For the parametrization  $F_2\approx x^\lambda$, the "effective power"
$\lambda$ rises on average from about 0.15 around
$Q^2\approx 1$\g2  to 0.4 at $Q^2\approx 1000$\g2 [3]. This
exponent cannot be identified with the intercept - 1 of a simple Pomeron
pole since by factorization it cannot depend on the virtuality of the
external particle. A $Q^2-$ dependent intercept, compatible with
the data, may arise from unitarization. However such a model
[4] leaves much flexibility since neither the input (Born) value of
the intercept is known for sure, nor a reliable unitarization
procedure exists (for a recent attempt see however [5]).
Moreover, claims exist that the HERA data are compatible with a
softer, namely logarithmic behavior in $x$ (obeying the Froissart 
bound) with a factorized $Q^2$ dependence [6, 7].

On the other hand the $Q^2$, or GLAP, evolution in the "leading-log"
approximation, has the following asymptotic solution for the
singlet SF, valid for low $x$ and high $Q^2$ [1, 8]
$$F_2\approx \sqrt{\gamma_1\ell n(1/x)\ \ell n\ell n{Q^2}} \ ,     \eqno(1.1)$$
with $\gamma_1={16N_c\over(11-2f/3)}$.  For 4 flavours $(f=4)$ and three
colours $(N_c=3)$, one gets $\gamma_1=5.76$.

The asymptotic solution of the BFKL evolution equation is the 
so-called "Lipatov Pomeron" [2]. The  numerical value of its 
intercept was calculated [2] to be between 1.3 and 1.5. This large 
value gave rise to speculations that the "Lipatov Pomeron" has been 
seen at HERA, where the large - $Q^2$ data seemed to be compatible 
with a steep rise  $\approx x^{-0.4}$ (for an alternative 
interpretation of the relation between the "Lipatov Pomeron" and the         
"HERA effect" see for example [9]). However, according to the results 
of a recent calculation [10], the sub-asymptotic 
corrections to the Pomeron pole in perturbative QCD are larger than 
expected and they contribute distructively to the intercept, thus 
lowering its value and making it compatible with the intercept of the soft
Pomeron $\footnote {One of us (F.P.) thanks B.I. Ermolaev for a 
discussion of this issue.}$.

The technical difficulties of the purely perturbative calculations
are aggravated by the unpredictable non-perturbative contributions,
both in the BFKL and GLAP evolutions, thus reducing the precision of
the theoretical calculations and their predictive power. All these
difficulties are redoubled by the unknown unitarity corrections to be
included in the final result.

Attempts to extrapolate the Pomeron-dominated "soft" SF by
applying GLAP evolution towards higher $Q^2$ are known in the
literature (see \eg [4, 11]). They differ in some details, namely
in the choice of the model for the Pomeron, its range, \ie  the
value of $Q^2$ from which the evolution starts, and in the details of
the evaluation (explicit in [11] or numerical in [4a]) of this
evolution. We are not aware of any results of an "inverse
extrapolation".

The situation has been recently summarized [3] in a figure (see also Sec. 6)
showing the $x$ - and $Q^2$ - dependence of the derivative       
$dF_2/d\ell n Q^2$.
The philosophy behind this figure is
that the turning point (located at $Q^2\sim2$ \g2 )
divides "soft" and "hard" dynamics. As shown
in [3], one of the most successful approaches to
the GLAP evolution, that by Ref. [12], fails to follow the soft dynamics.
A phenomenological model (called "ALLM") for the structure function and      
cross-section, applicable in a wide range of their kinematical
variables is well known in the literature [13].
Recently [14] it was updated to fit the data and shown to exhibit both -
the rising and falling - parts of the derivative versus $x$ (or $Q^2$).
We will comment more the behavior of this derivative in Sec. 6.

Below we pursue a pragmatic approach to the problem. We seek for an
interpolation formula between the known asymptotic solutions imposed as    
boundary conditions. Clearly, such an interpolation is not unique,           
but it seems to be among the simplest. Moreover, it fits the data
remarkably well, thus indicating that the interpolation is not far
from reality.

\bigskip

{\bf 2 KINEMATICS}

\medskip

We use the standard kinematic variables to describe deep
inelastic scattering:
$$ e(k) \ +\ p(P) \ \to\ e(k')\ +\ X\ ,                          \eqno(2.1)$$
where $ k, k', P$ are the  four-momenta of the incident electron, scattered
electron and incident proton.
$ Q^2$ is the negative squared four-momentum transfer carried by
the virtual exchanged boson (photon)
$$ Q^2 \ =\ -q^2\ =\ -(k-k')^2\ ,                                \eqno(2.2)$$
$x$ is the Bj\"orken variable
$$ x\ =\ {Q^2\over 2P.q} \ ,                                     \eqno(2.3)$$
$y$ (the inelasticity parameter) describes the energy transfer to the 
final hadronic state $$y\ =\ {q.P\over k.P}    \ ,                
                                                                 \eqno(2.4)$$ 
$W$ is the center of mass energy of the  $\gamma^* p$ 
system $$ W^2 \ =\ Q^2{1-x\over x}+m_p^2 \ ,                     \eqno(2.5)$$ 
with $m_p$, being the proton mass. Note that only two of these variables 
are independent and that, at  high energies for a virtual photon with 
$x\ll 1$, one has $W^2\ \sim {Q^2\over x}$.

\bigskip
                          
   {\bf 3 STRUCTURE FUNCTION FOR SMALL $x$ AND ALL $Q^2$}

\medskip

Following the strategy outlined in the Introduction, we suggest the
following ansatz for the small-$x$ singlet part (labelled by the
upper index $S,0$) of the proton structure function, interpolating
between the soft (VMD, Pomeron) and hard (GLAP evolution) regimes:

$$F_{2}^{(S,0)}(x,Q^2) =
A\left({Q^2\over Q^2+a}\right)^{1+\widetilde{\Delta} (Q^2)}
e^{\Delta (x,Q^2)}  ,                                             \eqno(3.1)$$
with the "effective power"
$$ \widetilde{\Delta }(Q^2) =\epsilon+\gamma_1\ell n {
\left(1+\gamma_2\ell n{\left[1+{Q^2\over Q^2_0}\right]}\right)} ,
                                                                  \eqno(3.2)$$
and
$$\Delta (x,Q^2) = \left(\widetilde{\Delta } (Q^2)  \ell n{x_0\over x}\right)
                   ^{f(Q^2)},                                     \eqno(3.3)$$
where
$$ f(Q^2) = {1\over 2}\left( {1+e^{-{Q^2/Q_1^2}}}\right) .        \eqno(3.4)$$

At small and moderate values of $Q^2$ (to be specified from the fits, see
below), the exponent $\widetilde{\Delta}(Q^2)$ (3.2) may be interpreted
as a $Q^2$-dependent "effective Pomeron intercept".

The function $f(Q^2)$ has been introduced in order to provide for the
transition from the Regge behavior, where $f(Q^2)=1$, to the 
asymptotic solution of the GLAP evolution equation, where 
$f(Q^2)=1/2$.

\smallskip

By construction, the model has the following asymptotic limits:

\smallskip

a) Large $Q^2$, fixed $x$:
$$ F_{2}^{(S,0)}(x,Q^2\to \infty)\to A\
\exp^{\sqrt{\gamma_1\ell n\ell n{Q^2\over Q_0^2}\ \ell n{x_0\over x}}}\ ,
                                                                    \eqno(3.5)$$
which is the asymptotic solution of the GLAP evolution equation
(see Sec. 1).

\smallskip

b) Low $Q^2$, fixed $x$:
$$F_{2}^{(S,0)}(x,Q^2\to 0) \to A\
e^{\Delta (x,Q^2\to 0)} \
\left({Q^2\over a}\right)^{1+\widetilde{\Delta }(Q^2\to 0)}       \eqno(3.6)$$
with $$ \widetilde{\Delta }(Q^2\to 0) \to
\epsilon+\gamma_1 \gamma_2   { \left({{Q^2\over Q^2_0} }\right)}\
\to\ \epsilon ,                                                   \eqno(3.7)$$
$$ f(Q^2\to 0) \to 1 ,                                            \eqno(3.8)$$
whence
$$F_{2}^{(S,0)}(x,Q^2\to 0) \to A\ \left( {x_0\over x}
\right)^\epsilon \ \left({Q^2\over a}\right)^{1+\epsilon} \ \propto
(Q^2)^{1+\epsilon} \ \to 0\ ,                                     \eqno(3.9)$$
as required by gauge invariance.

\smallskip

c) Low $x$, fixed $Q^2$:

$$F_{2}^{(S,0)}(x\to 0,Q^2) \ =\
A\left({Q^2\over Q^2 + a}\right)^{1+\widetilde{\Delta }(Q^2)}
e^{\Delta (x\to 0,Q^2)}  .                                        \eqno(3.10)$$
If
$$f(Q^2)\sim 1\ ,                                                 \eqno(3.11)$$
\ie when $Q^2\ll Q_1^2$,
we get the standard (Pomeron-dominated) Regge behavior (with a $Q^2$
dependence in the effective Pomeron intercept)
$$F_{2}^{(S,0)}(x\to 0,Q^2) \to A\ \left({Q^2\over Q^2 +
a}\right)^{1+\widetilde{\Delta}(Q^2)}\ \left({x_0\over x}
\right)^{\widetilde{\Delta}(Q^2)} \ \propto
x^{-\widetilde{\Delta}(Q^2)}  .
                                                                  \eqno(3.12)$$

Within this approximation, the total
cross-section for $(\gamma,p)$ scattering as a function of the center
of mass energy $W$ is
$$ \sigma ^{tot,(0)}_{\gamma,p} (W)=
4\pi^2\alpha\ \left[{F_{2}^{(S,0)}(x,Q^2)\over
Q^2}\right]_{Q^2\to 0}  =\ 4\pi^2\alpha\ A\  a^{-1-\epsilon}\
x_0^\epsilon\ W^{2\epsilon} .                                     \eqno(3.13)$$

\bigskip

   {\bf 4 EXTENSION TO LARGE $x$}

\medskip

In this section we complete our model by including the large-$x$
domain, extending to $x=1$, and for all kinematically allowed $Q^2$.
Since we are essentially concerned with the
small-$x$ dynamics (transition between the GLAP and BFKL evolution),
the present extension serves merely to have as good fits as possible
with a minimal number of extra parameters. To this
end we rely on the existing successful phenomenological models,
in particular on that of [4a] (CKMT).

Following CKMT, we multiply the singlet part of the above structure function
$F_{2}^{(S,0)}$
(defined in (3.1-3.4)) by a standard large-$x$ factor to get
$$F_{2}^{(S)}(x,Q^2) = F_{2}^{(S,0)}(x,Q^2)\
(1-x)^{n(Q^2)},                                                  \eqno(4.1)$$
with
$$ n(Q^2) ={3\over 2} \left(1+{Q^2\over Q^2+c}\right),           \eqno(4.2)$$
where $c=3.5489 $ GeV$^2$ [4a].

Next we add the nonsinglet $(NS)$ part of the structure function,
also borrowed from CKMT
$$F_{2}^{(NS)}(x,Q^2) =
B\ (1-x)^{n(Q^2)}\ x^{1-\alpha_r}\ \left({Q^2\over Q^2+b}\right)^{\alpha_r} \,
.                                                                 \eqno(4.3)$$
The free parameters that appear with this addendum are $c, B, b$ and
$\alpha_r$.  The final and complete expression for the proton
structure function thus becomes
$$F_{2}(x,Q^2) = F_{2}^{(S)}(x,Q^2) + F_{2}^{(NS)}(x,Q^2)\ .      \eqno(4.4)$$

The total cross-section for $(\gamma,p)$ scattering is
$$ \sigma ^{tot}_{(\gamma,p)} (W)=\
4\pi^2\alpha\ \left( A\  a^{-1-\epsilon}\ x_0^\epsilon\
W^{2\epsilon} + B\ b^{-\alpha_r}\  W^{2(\alpha_r-1)}\ \right) \ .
                                                                  \eqno(4.5)$$

\bigskip

  {\bf 5 FITTING TO THE DATA}

\medskip

In fitting to the data, the complete experimental "H1" data set
(which encloses 237 points:  193 from [15] and 44 from [16]) for the
proton structure function $F_{2}(x,Q^2)$ was used as well as, 76 data
points [17] on the $(\gamma,p)$  total cross-section
$ \sigma ^{tot}_{(\gamma,p)} (W)$.

We note that among a                                                        
total of 12 parameters, 8 are free, the resting 4 being fixed in the
following way:

1. $\epsilon=0.08$ is a "canonical" value [18], leaving
little room for variations (although, in principle, it can be also
subject to the fitting procedure);

2. when left free in the fitting procedure, $x_0$ takes a value
slightly beyond 1. Thus, we can safely fix $x_0=1$ without
practically affecting the resulting fits;

3. as already mentioned, we have set $c=3.5489$ GeV$^2$ relying on CKMT.
This parameter is responsible for the large-$x$ and
small-$Q^2$ region, outside the domain of our present interest;
                                                                              
4. as argued above, we may estimate from QCD the parameter
$\gamma_1=16N_c(11-2f/3)$ with four flavours $(f=4)$ and
three colors $(N_c=3)$, it equals 5.76. It corresponds to the
asymptotic regime (when $Q^2\rightarrow\infty$, or $f(Q^2)\rightarrow 1/2))$,
far away from the region of the fits, where  $f=1$ is
more appropriate, hence the value ${\gamma_1}=\sqrt{5.76}=2.4$ 
is more appropriate in the domain under consideration. Remarkably, 
this value comes also independently from the fits if $\gamma_1$ is 
let free.

To compare with, the CKMT model [4a] depends on 8 adjustable
parameters in the "soft" region, to be completed by QCD
evolution at higher values of $Q^2$, and with a higher twist term
added.
On the other hand, the proton structure function and $(\gamma^*,p)$ cross
section in the ALLM model [13,14] are given explicitly in the 
whole range of the kinematical variables,  and the fits to the data 
are good with a total of 23 adjustable parameters.

When limiting the fitted data to the structure function only [15,16] 
with $x<0.1$ (all $Q^2$), the singlet contribution alone, as 
approximated in Sec. 3, gives a very good fit (\xi2 $\sim 0.59$), 
shown in Figs. 1a, 2a.  We mention that this result is obtained with 
an economical set of 8 parameters (5 free), listed in Table 1.
                                                                              
The complete model of Sec. 4 gives very good fits in the whole ranges in $x$,
$Q^2$ and $W$ covered by measurements.
To be specific, we find \xi2 $\sim$ 0.69. We show the contributions to the
$\chi^2$ of the 3 data sets we used in Table 2, the numerical values for the
12 parameters (8 free) are presented in Table 3.

The results of our fits for the structure function versus
$Q^2$ for fixed $x$ are shown in Fig. 1 b and for fixed $Q^2$ as a
function of $x$ are in Figs. 2b, 3. The total cross section for
real photons on protons as function of $W$ is displayed in Fig. 4.

\bigskip

  {\bf 6  INTERFACE BETWEEN SOFT AND HARD DYNAMICS AND TRANSITION 
FROM BFKL TO GLAP EVOLUTION}

\medskip

{\bf 6.1 ${\partial F_{2}\over \partial (\ell n Q^2) }$ as a function of $x$ 
and $Q^2$.}
\medskip

The derivative of the SF with respect to $\ell n Q^2$ (slope for brevity)
measures
the amount of the scaling violation and eventually shows the
transition from soft to hard dynamics. This derivative depends on two
variables ($x$ and $Q^2$).
It was recently calculated from the HERA data [3]; in Figs. 5, 6a
we have quoted the corresponding results. In those
calculations the variables $x$ and $Q^2$ are strongly correlated,
it is implied that, for a limited acceptance (as it is the case       
in the HERA experiments) and for a fixed energy, one always
has a limited band in $Q^2$ at any given $x$, with average $Q^2$
becoming smaller for smaller $x$. From a theoretical point of view,
however, $x$ and $Q^2$ are quite independent and one is not restricted to
follow a particular path on the surface representing the slope.
Therefore we plot in Fig. 5 the slope calculated from our model 
(4.4) with the parameters fitted to the data, in one more dimension than usual,
\ie as a function of the two independent variables - $x$ and $Q^2$.
The two slopes on the hill of ${\partial F_{2}\over \partial (\ell n Q^2) }$
in Fig. 5  correspond to soft and hard dynamics. The
division line is only symbolic since there is a wide interface region
where both dynamics mix, each tending to dominate on the lower side
of its own slope.
Remarkably, the division line - or line of maxima of this surface - turns out
to be almost $Q^2$-independent ($\sim 40$ \g2 ).
The difference with the maximum at 2 \g2 exhibited in [3] is due to the 
special experimental set of ($x,Q^2$) chosen in [3], discussed above and shown
in Fig. 5.

Notice that the slope becomes negative in a region between $Q^2\sim 
200$ and $\sim 4000$~\g2 , at small $x$~; this region tends to 
narrow when $x$ increases beyond $x=$ 0.0005
and finally disappears  when $x$ exceeds 0.05.
         
The same results are exposed on
families of 2-dimensional figures as well (Figs. 6a, 6b) showing the 
$x$ - (and $Q^2$ -) dependence of the slope when the other variable 
takes fixed values.  Fig. 6a shows that our predictions are quite in 
agreement with the data from [3]; also shown is the failure of the 
approach of the GLAP evolution equation [12] to 
follow the low $x$ ($Q^2$) dynamics as reported in [3].  
Fig. 6b shows the variation 
with $x$ of the region with negative slope.  Notice that the rising 
part to large extent is a threshold effect due to the increasing 
phase space (see [19]).

\medskip

{\bf 6.2 ${\partial \ell n F_{2}\over \partial (\ell n (1/x))} $ as a function of
$Q^2$ for some $x$ values.}

\medskip

The derivative of the logarithm of the SF with respect to $\ell n 1/x$,
when measured in the Regge region,
can be related (for low $x$) to the Pomeron intercept. In Fig. 7
the $Q^2$-dependence of this derivative is shown for some low $x$ - values,
together with the "effective
power" $\tilde\Delta$ (3.2). On the same figure, the behavior of the
function $f(Q^2)$ (3.4) is also shown. In our model, Regge                   
behavior is equivalent to the condition that $f(Q^2)$ is close to
unity. This lower limit, marked on Fig. 7 (tentatively approximated           
within a 2 \% accuracy for the function $f(Q^2)$), is located near 
$40$~\g2 .  Until this landmark, the effective power $\tilde\Delta$ indeed 
remains very close to ${\partial \ell n F_{2}\over \partial (\ell n (1/x))} $, 
beyond Regge behavior is not valid (since $f\ \ne 1$) and $\tilde\Delta$ 
cannot be considered as the effective slope any more. On the other 
hand,  ${\partial \ell n F_{2}\over \partial (\ell n (1/x))} $
turns down as $Q^2$ increases, approaching 
its "initial value" of $\approx 0.1$ at largest $Q^2$ and coming 
closer to the unitarity bound. Notably, at large $Q^2$ the derivative 
gets smaller as $x$ decreases, contrary to the general belief 
that dynamics becomes harder for smaller $x$, but in accord with an 
observation made in [20]. Care should be however taken in 
interpreting the "hardness" of the effective power outside the Regge 
region.

According to our model, the change from the BFKL (Pomeron) to the
GLAP evolution occurs when $f(Q^2)$ changes from 1 to 1/2. This
variation happens in a band in $Q^2$, namely
between $\sim 40$ \g2 and $\sim 4000$ \g2 .

Let us remind once more that our interpolating formula (3.1)
between Regge behavior and GLAP evolution was suggested for
small $x (\ x\leq0.1)$. The larger $x$ part was introduced for
completeness and better fits only, without any care of its
correspondence to the GLAP evolution equation. It does not affect
however the kinematical domain of the present and future HERA
measurements and Pomeron dominance (BFKL evolution) we are interested
in.

\bigskip

  {\bf 7 CONCLUSIONS}

\medskip

Once the "boundary conditions" (at low and high $Q^2$) are satisfied,
the interpolation may be considered as an approximate solution valid
for all $Q^2$. Clearly, our interpolation is not unique. For
example, the choice of $f(Q^2)$, satisfying the boundary conditions
$f(0)=1$ and $f(\infty)=1/2$, may be different from ours. However,
there is little freedom in the choice of the asymptotic forms,
different from those we have used, namely (3.5) and (3.12). The
utilization of a soft Pomeron input different from (3.12) is
credible. For example, a dipole Pomeron was shown [7, 11] to have the
required formal properties and to fit the data at small and moderate
$Q^2$.  Moreover, the dipole Pomeron does not violate the Froissart
bound, so it does not need to be unitarized. Attempts [6, 7] to fit
the high-$Q^2$ HERA data without a power in $x$, i.e. with
logarithmic functions, attributing the whole $Q^2$ dependence to the
(factorized) "residue function", are disputable. What is even more     
important from the point of view of the present interpolation, a
power in $x$ must be introduced anyway to match the high $Q^2$ GLAP
evolution solution (3.5).  This discussion brings us back to
the interesting but complicated problem of unitarity.

As it is well known,
the power increase of the total cross sections, or of
the SF towards small $x$ cannot continue indefinitely.
It will be slowed down by unitarity, or shadowing corrections, whose
calculation or even recipe - especially for high virtualities $Q^2$ -
is a delicate and complicated problem, beyond the scope of the present
paper. 
Here we only mention, that once the model fits the data, it cannot be far from
the "unitarized" one in the fitted range, since the data "obey"
unitarity.

To conclude:

1. Strong interaction dynamics is continuous, hence the relevant
solutions should be described by continuous solutions as well;

2. The formal solutions of the GLAP equations, even in their most
advanced forms, ultimately contain some freedom (e.g. "higher
twists", or non-perturbative corrections) or approximations;

3. However so elaborated or "precise" the existing solutions are,
unitarity corrections will modify their form anyway;

The above remarks justify the use for practical purposes of an
explicit solution that satisfies the formal theoretical requirements
and yet fits the data. Its simplicity and flexibility make possible its 
further improvement and its use as a laboratory in studying 
complicated and yet little understood transition phenomena.

\bigskip
{\bf ACKNOWLEDGEMENTS}

We thank M. Bertini for discussions and L.A. Bauerdick for a useful 
correspondence.

    \vfill\eject

\centerline{\bf References}
\smallskip

[1] V. N. Gribov, L. N. Lipatov, Sov. J. Nucl. Phys. {\bf 15}, 
438 and 675  (1972); 
G. Altarelli, G. Parisi, Nucl. Phys. B {\bf 126}, 298 (1977).

[2] Y. Y. Balitskii, L. N. Lipatov, Sov. Physics JETP
{\bf 28}, 822 (1978); E. A. Kuraev, L. N. Lipatov, V. S. Fadin,
ibid. {\bf 45}, 199 (1977); L. N. Lipatov, ibid. {\bf 63}, 904 (1986).

[3] L. Bauerdick, {\it Proton structure function and ($\gamma^*,p$) cross
section at HERA} in  {\bf Interplay between Hard and
Soft Interactions in Deep Inelastic Scattering, Max Planck workshop, 
Heidelberg, 1997} (transparencies available from: http://www.mpi-hd.
mpg.de/~hd97/).

[4] a) A. Capella, A. Kaidalov, C. Merino, J. Tran Thanh Van,
   Phys. Lett. B {\bf 337}, 358 (1994).

    b) M. Bertini, P. Desgrolard, M. Giffon, E. Predazzi,
   Phys. Lett. B {\bf 349}, 561 (1995).

[5] A. Capella, A. Kaidalov, V. Neichitailo, J. Tran Thanh Van,
LPTHE-ORSAY 97-58 and hep-ph/9712327, 1997.

[6] W. Buchmuller, D. Haidt, DESY/96-61, 1996.

[7] L. L. Jenkovszky, E. S. Martynov, F. Paccanoni, 
{\it Regge behavior of the nucleon structure functions}, PFPD 95/TH/21,
1995; 
L. L. Jenkovszky, A. Lengyel and F. Paccanoni, {\it                             
Parametrizing the proton structure function}, hep-ph/9802316, 1998.

[8] R. D. Ball, S. Forte. Phys. Lett. B {\bf 335}, 77 (1994); 
F. Paccanoni, {\it Note on the DGLAP evolution equation}, in {\bf Strong 
Interaction at Long Distances}, edited by L. L. Jenkovszky (Hadronic 
Press, Palm Harbor, 1995).

[9] M. Bertini, M. Giffon, L. L. Jenkovszky, F.Paccanoni,  
E. Predazzi, Rivista Nuovo Cim. {\bf 19}, 1 (1996).

[10] V. S. Fadin, L. N. Lipatov, {\it BFKL Pomeron in the next-to-leading
approximation}, \hfill\break
hep-ph/9802290, 1998.

[11] L. L. Jenkovszky, A. V. Kotikov, F. Paccanoni, Phys. Lett. B  
{\bf 314}, 421 (1993).

[12] M. Gl\"uck, E. Reya, A. Vogt. Zeit. Phys. C {\bf 67}, 433 (1995).

[13] H. Abramowicz, E. Levin, A. Levy, U. Maor, Phys. Lett. B
{\bf 269}, 465 (1991).

[14]  H. Abramowicz, A. Levy, DESY 97-251 and hep-ph/9712415, 1997.

[15] S. Aid \etal H1 collaboration,  Nucl. Phys. B {\bf 470}, 3 (1996):

193 values of the structure function of the proton, for $Q^2$,
between 1.5\g2 and 5000\g2, and $x$, between 3 10$^{-5}$ to 0.32
(of which 169 data correspond to $x < 0.1$).

[16] C. Adloff \etal H1 collaboration, Nucl. Phys. B {\bf 497}, 3 (1997):

44 values of the structure function of the proton, at low $x$,
down to 6 10$^{-6}$, and low $Q^2$, between 0.35\g2 and 3.5\g2.

[17] D.O. Caldwell \etal, Phys. Rev. D {\bf 7}, 1384  (1975)~;
Phys. Rev. Lett. {\bf 40}, 1222 (1978).

M. Derrick \etal ZEUS collaboration, Zeit. Phys. C {\bf 63}, 391 (1994).
                                                                               
S. Aid \etal H1 collaboration, Zeit. Phys. C {\bf 69}, 27 (1995).

[18] H.Cheng, J.K. Walker, T.T. Wu, Phys. Lett. B {\bf 44}, 97 (1973)~;

A. Donnachie, P.V. Landshoff, Phys. Lett. B {\bf 296}, 493 (1975).

[19] L. L. Jenkovszky, E. S. Martynov, F. Paccanoni, Nuovo Cimento A 
{\bf 110}, 649 (1997).

[20] A. De Roeck, E. A. De Wolf, Phys. Lett. B {\bf 388}, 188 (1996).

                         \vfill\eject
\vglue 2.cm
\centerline{\bf Tables captions}

\bigskip

{\bf Table 1.}

Parameters used in our "first approximation fit" ($x<0.1$).

\bigskip

{\bf Table 2.}

$\chi^2$ - contributions of each set of data used in our fit with the parameters
listed in Table 3.

\bigskip
{\bf Table 3.}

Parameters used in our fit in the whole kinematical range (see the text).

                                 \vfill\eject
\vglue 1.cm
\centerline{\bf Figures captions}
\medskip

{\bf Fig. 1 a}
Proton structure function $F_{2}(x,Q^2)$ as a function of $Q^2$ at
various values of fixed $x$. For a better display, the structure function
values have been scaled at each $x$ by the factor shown in brackets on the
same line as the $x$ values. The shown H1 - data are from [15,16],
the error bars represent the statistical and systematic errors added in
quadrature, the curves are the results of our first parametrization fitted on
$x<0.1$ data ("low-$x$, Pomeron dominated" approximation,
the parameters being listed in Table 1).

\medskip

{\bf Fig. 1 b}
Proton structure function $F_{2}(x,Q^2)$ as a function of $Q^2$ at
various values of fixed $x$ as in Fig. 1 a but the curves being the results
of our second parametrization fitted to all H1 data
[15,16] of the proton structure function and to the total
cross-sections  of the $(\gamma ,p)$ process [17]
(the parameters are listed in Table 3).

\medskip

{\bf Fig. 2 a}
Proton structure function $F_{2}(x,Q^2)$ as a function of $x$ at
various values of fixed $Q^2$. Results of our first approximation, see also
Fig. 1 a.

\medskip

{\bf Fig. 2 b}
Proton structure function $F_{2}(x,Q^2)$ as a function of $x$ at
various values of fixed $Q^2$. Results of our second parametrization, see also
Fig. 1 b.

\medskip

{\bf Fig. 3}
Proton structure function $F_{2}(x,Q^2)$ as a function of $x$ at
various low $Q^2$ values. See also Fig. 1 b.

\medskip

{\bf Fig. 4} Total cross-section of the reaction $(\gamma,p)$
$ \sigma ^{tot}_{(\gamma,p)} $ as a function of $W$, center of mass energy.
(see also Fig. 1 b).

\medskip

{\bf Fig. 5}
Two-dimensional projection of the three dimensional
"slope" of the proton structure function. The surface represents
${\partial F_{2}(x,Q^2) \over \partial (\ln Q^2)}\ $ as a function of $x$ and
$Q^2$  as following from the present parametrization with its line of maximum
(open squares). The crosses are the points calculated from the HERA data in [3],
located on an experimental ($x,Q^2$) path.

\medskip

{\bf Fig. 6a}
Derivative of the proton structure function
${\partial  F_{2}(x,Q^2) \over \partial (\ln Q^2)} $ as a function of $x$, for
some $Q^2$ values as indicated. The round dots are the HERA data, 
the open squares the results from [12] taken from [3]
and the hollow triangles are the results of the present parametrization.

\medskip

{\bf Fig. 6b}
Same derivative as in Fig. 6a
${\partial F_{2}(x,Q^2)\over \partial (\ln Q^2)}\ $ as a function of $Q^2$, for
some $x$ values as indicated. The solid curves are the results of the present
parametrization.

\medskip

{\bf Fig. 7}
Derivative of the logarithm of the proton structure function
${\partial \ell n F_{2}(x,Q^2)\over \partial (\ell n (1/x))} $ versus $Q^2$ for
some $x$ values as indicated.
Also plotted on the same (left) scale is the
effective exponent $\widetilde{\Delta}$ (3.2), representing
the Pomeron intercept $-1.$ only when $f(Q^2)\approx 1$.
The function $f(Q^2)$ (3.4) is also shown as a dashed line (right scale);
the transition between the Regge behavior ($ f=1.$) and the GLAP
evolution ($f=0.5$) occurs within an estimated band located between vertical 
landmarks (see the text).

                         \vfill\eject

\def\init{\tabskip 0pt\offinterlineskip}
\def\crr{\cr\noalign{\hrule}}
$$\vbox{\init\halign to 12 cm{
\strut#&\vrule#\tabskip=1em plus 2em&
\hfil$#$&
\vrule#&
\hfil$#$\hfil&
\vrule#\tabskip 0pt\crr
&&{\bf A} &&0.1612&\cr
&&{\bf a{\rm \ (GeV)}^2} &&0.2133&\cr
&&{\bf  \gamma_2  }&&0.02086         &\cr
&&{\bf  Q_0^2 {\rm \ (GeV)}^2} &&0.2502         &\cr
&&{\bf  Q_1^2 {\rm \ (GeV)}^2} &&676.9         &\crr
&&{\bf  x_0  }&&1.0 \ {\rm (fixed)}         &\cr
&&{\bf \epsilon}&&0.08\ {(\rm fixed\ [18])}         &\cr
&&{\bf  \gamma_1  }&&2.4   \ {\rm (fixed\ QCD)}   &\crr}}$$

\centerline {\bf Table 1.}

Parameters used in our "first approximation fit" ($x<0.1$).

\bigskip

\def\init{\tabskip 0pt\offinterlineskip}
\def\crr{\cr\noalign{\hrule}}

\def\init{\tabskip 0pt\offinterlineskip}
\def\crr{\cr\noalign{\hrule}}
$$\vbox{\init\halign to 12 cm{
\strut#&\vrule#\tabskip=1em plus 2em&
\hfil$#$&
\vrule$\,$\vrule#&
\hfil$#$\hfil&
\vrule#&
\hfil$#$\hfil&
\vrule#\tabskip 0pt\crr
&&{\rm Data\ set} && {\rm N.\ of\ points}&& \chi^2&\crr
&& \sigma ^{tot}_{(\gamma,p)}\ (W>3\ GeV^2)\ {\rm  \ [17] } &&73 &&73 &\cr
&&F_{2},\ {\rm  H1\ [15] } &&193 &&116 &\cr
&&F_{2},\ {\rm  H1\ (low }\ x {\rm )\ [16] } &&44 &&20 &\crr
}}$$

\centerline {\bf Table 2.}
\medskip
$\chi^2$ - contributions of each set of data used in our fit with the parameters
listed in Table 3.

\bigskip

\def\init{\tabskip 0pt\offinterlineskip}
\def\crr{\cr\noalign{\hrule}}
$$\vbox{\init\halign to 12 cm{
\strut#&\vrule#\tabskip=1em plus 2em&

\hfil$#$&
\vrule#&
\hfil$#$\hfil&
\vrule#\tabskip 0pt\crr
&&{\bf A} &&0.1623&\cr
&&{\bf a{\rm \ (GeV)}^2} &&0.2919&\cr
&&{\bf  \gamma_2  }&&0.01936         &\cr
&&{\bf  Q_0^2 {\rm \ (GeV)}^2} &&0.1887         &\cr
&&{\bf  Q_1^2 {\rm \ (GeV)}^2} &&916.1         &\cr
&&{\bf B} &&0.3079&\cr
&&{\bf b{\rm \ (GeV)}^2} &&0.06716&\cr
&&{\bf \alpha_r}&&0.5135\ &\crr
&&{\bf  x_0  }&&1.0 \ {\rm (fixed)}         &\cr
&&{\bf \epsilon}&&0.08\ {\rm (fixed\ [18])}     &\cr                           
&&{\bf  \gamma_1  }&&2.4   {\rm (fixed\ QCD)}  &\cr
&&{\bf c{\rm \ (GeV)}^2} &&3.549\ {\rm (fixed\ [4a])}  &\crr
}}$$

\centerline {\bf Table 3.}

Parameters used in our fit in the whole kinematical range (see the text).

                
\begin{center}
\leavevmode\epsfbox{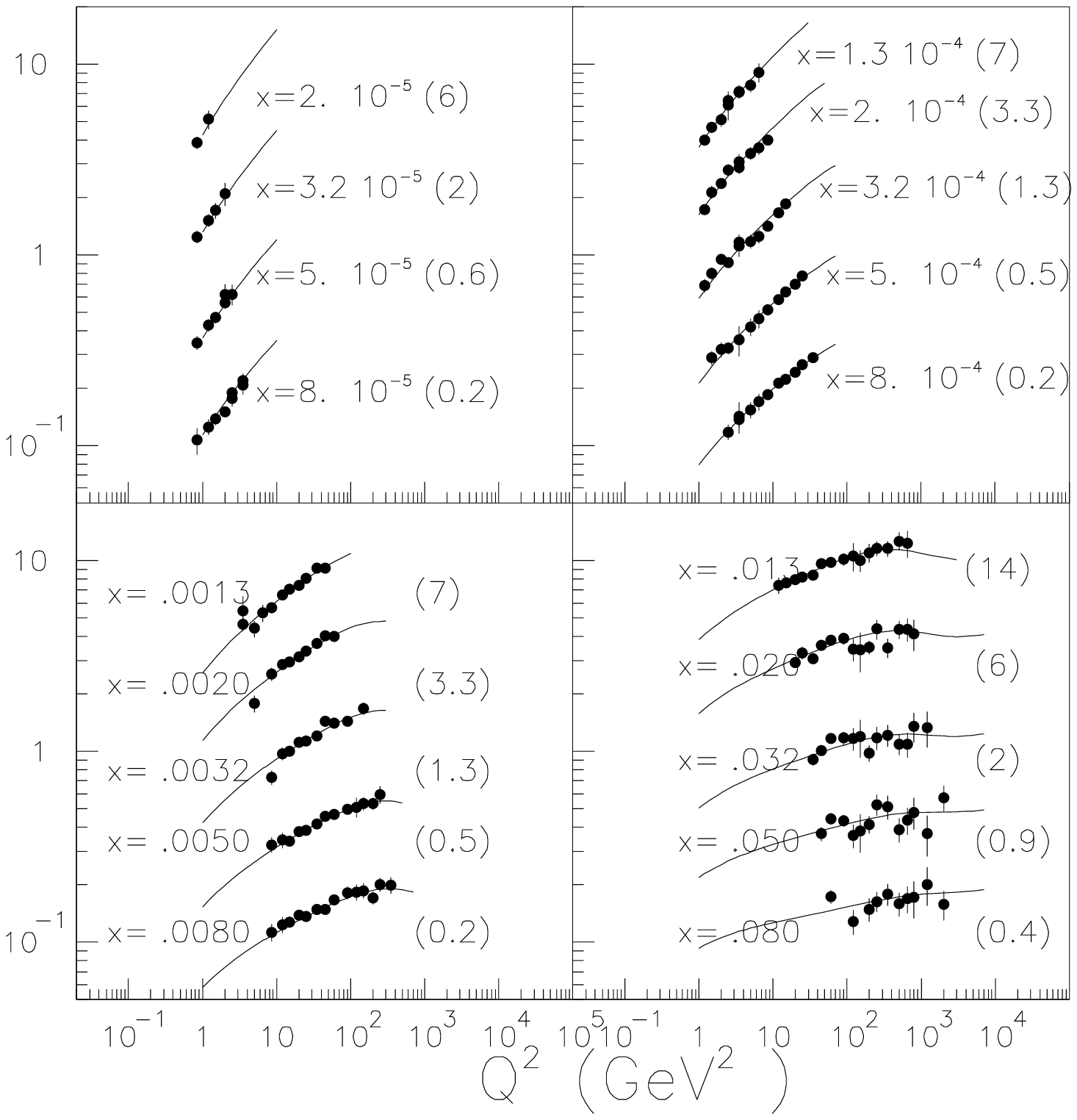} 

Fig. 1a
\end{center}

\begin{center}
\leavevmode\epsfbox{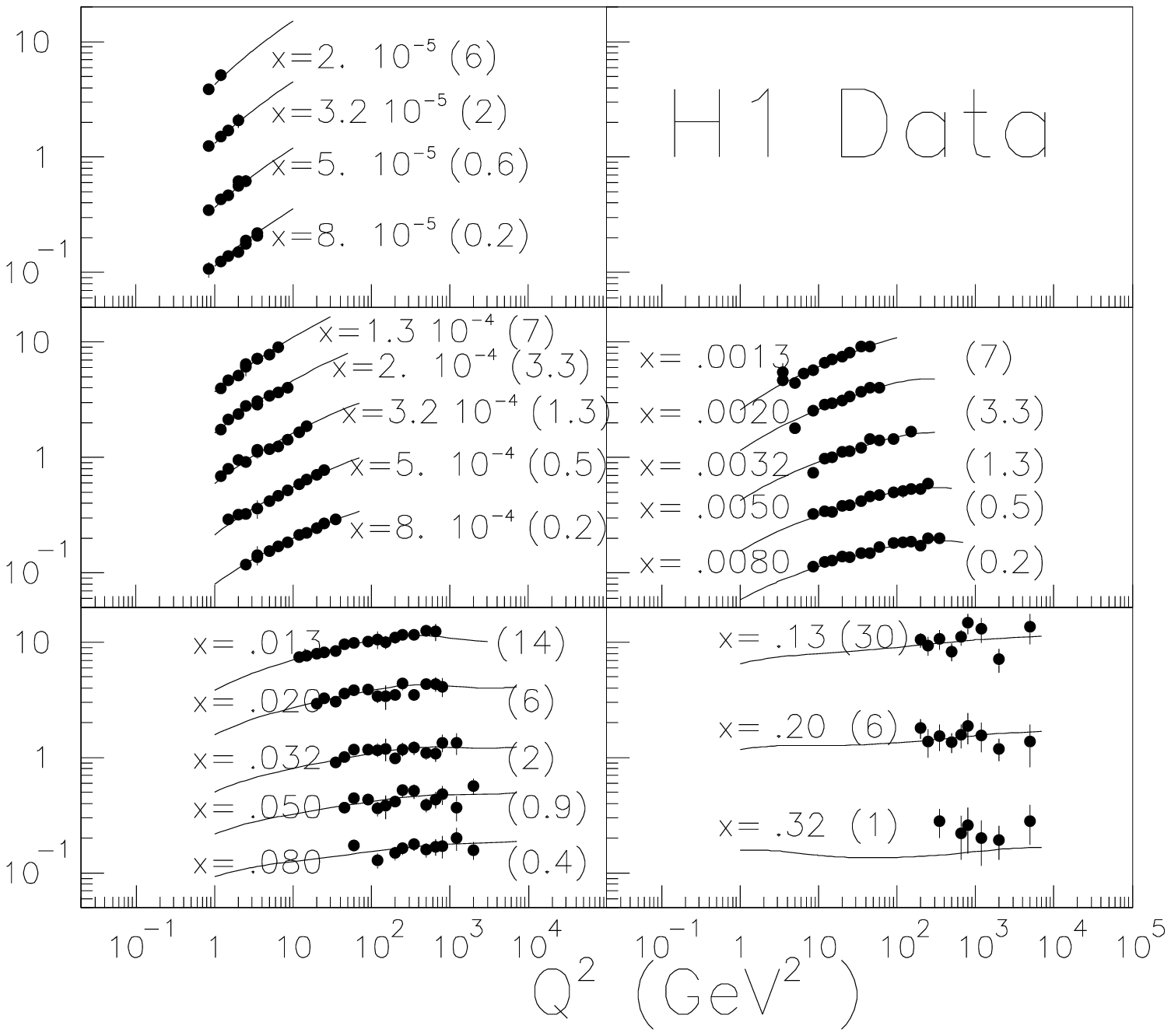}

Fig. 1b 
\end{center}

\begin{center}
\leavevmode\epsfbox{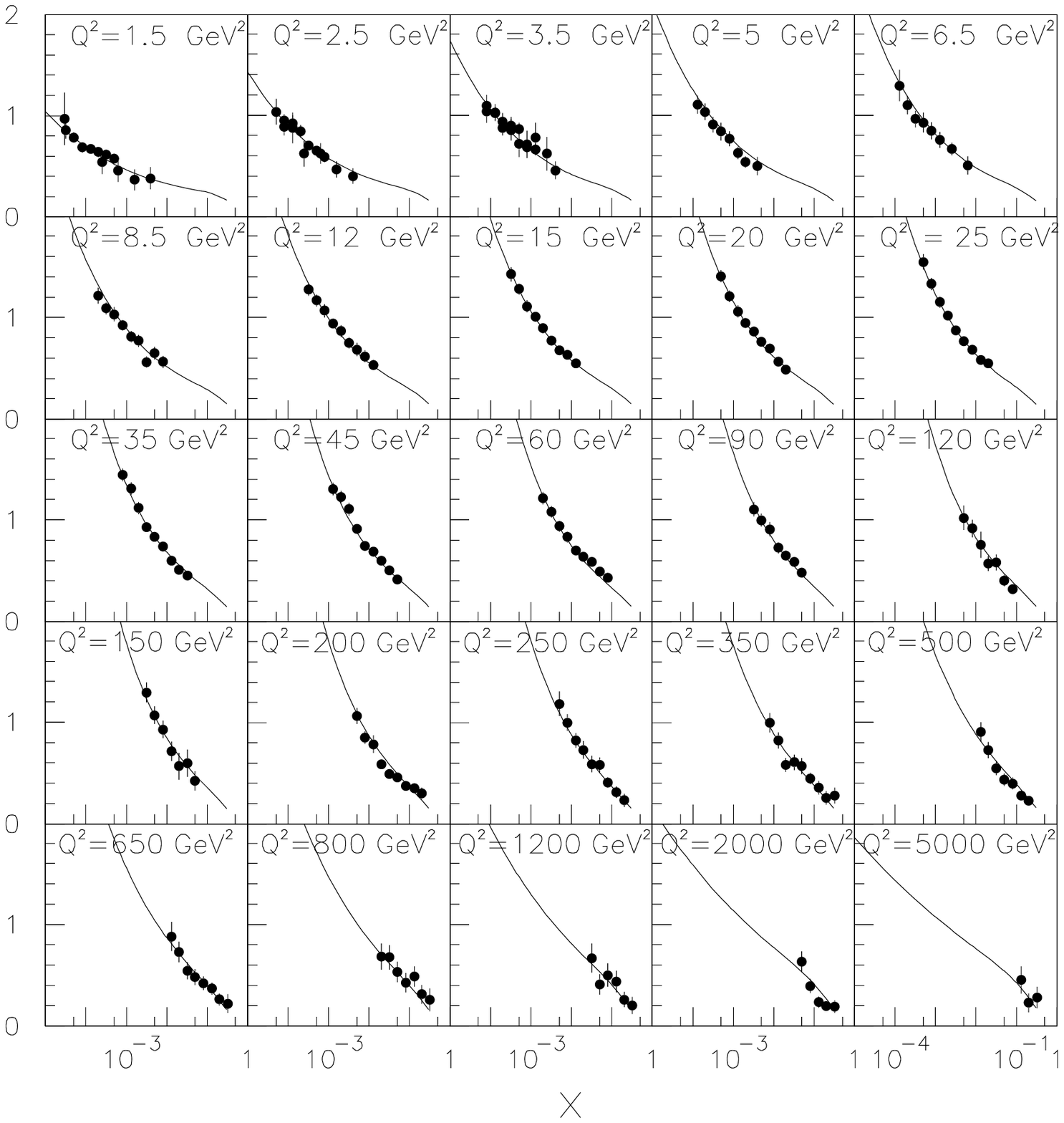}

Fig. 2a
\end{center}
                
\begin{center}
\leavevmode\epsfbox{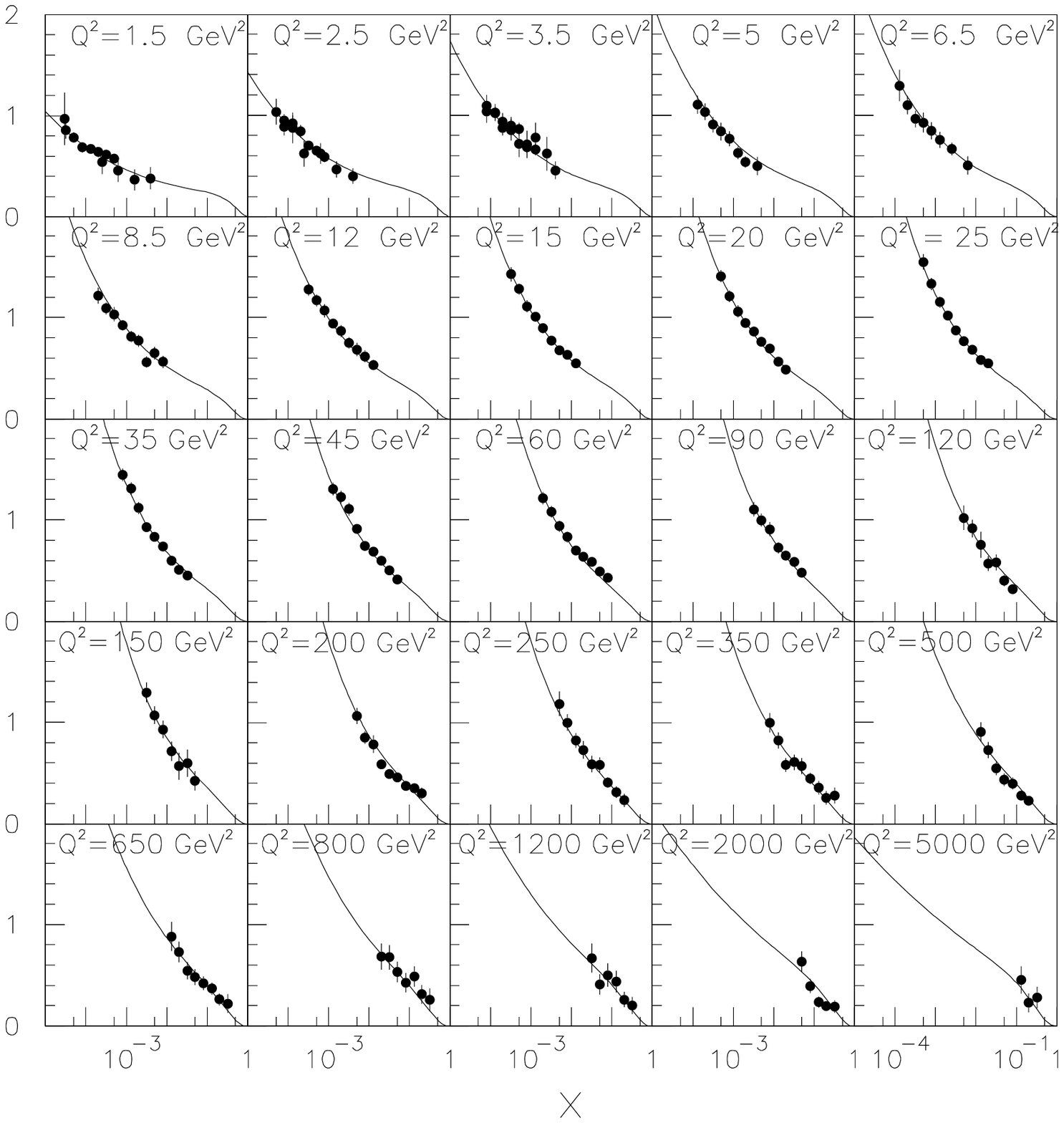}

Fig. 2b
\end{center}
                
\begin{center}
\leavevmode\epsfbox{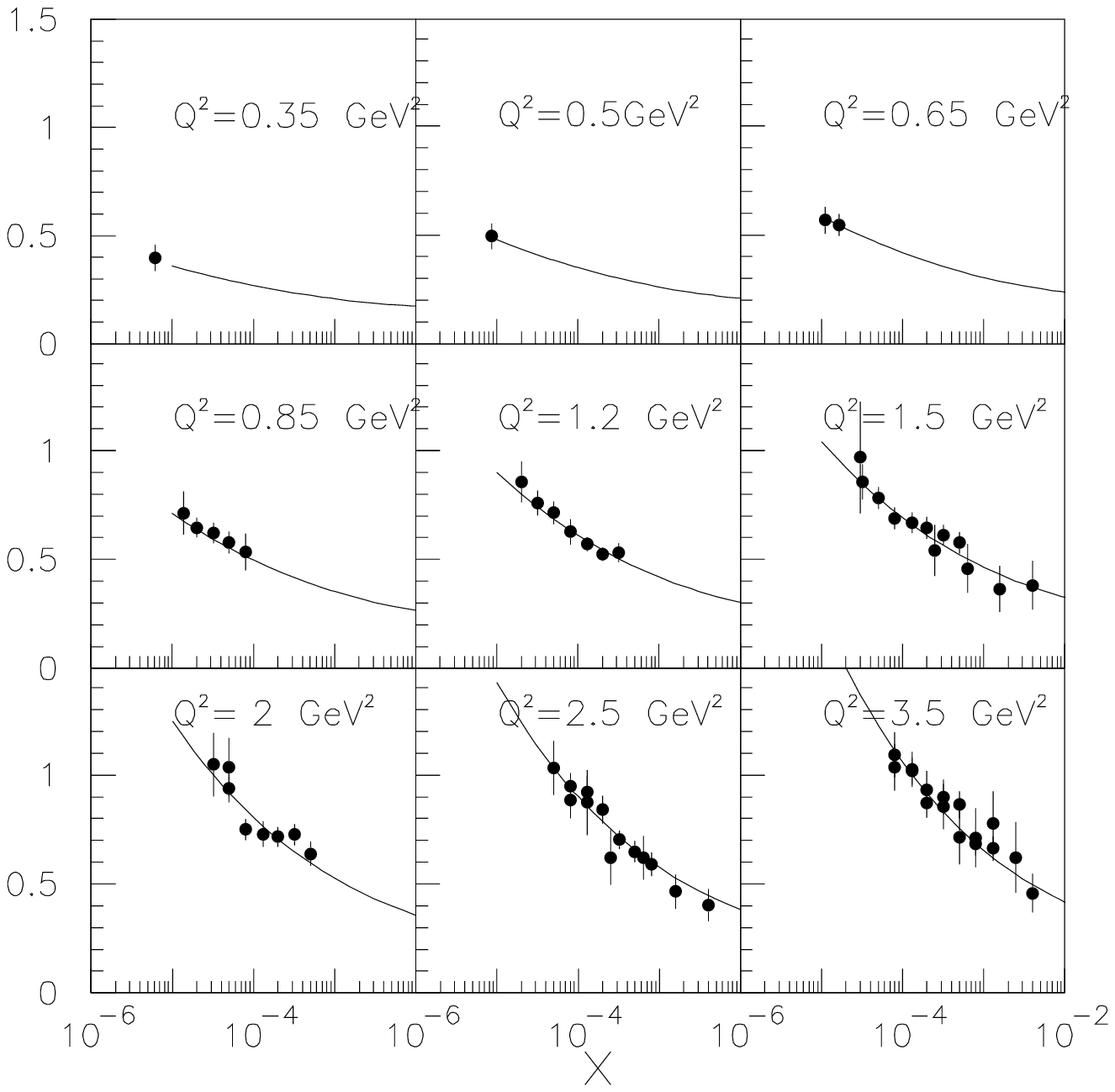}

Fig. 3
\end{center} 
\begin{center}
\leavevmode\epsfbox{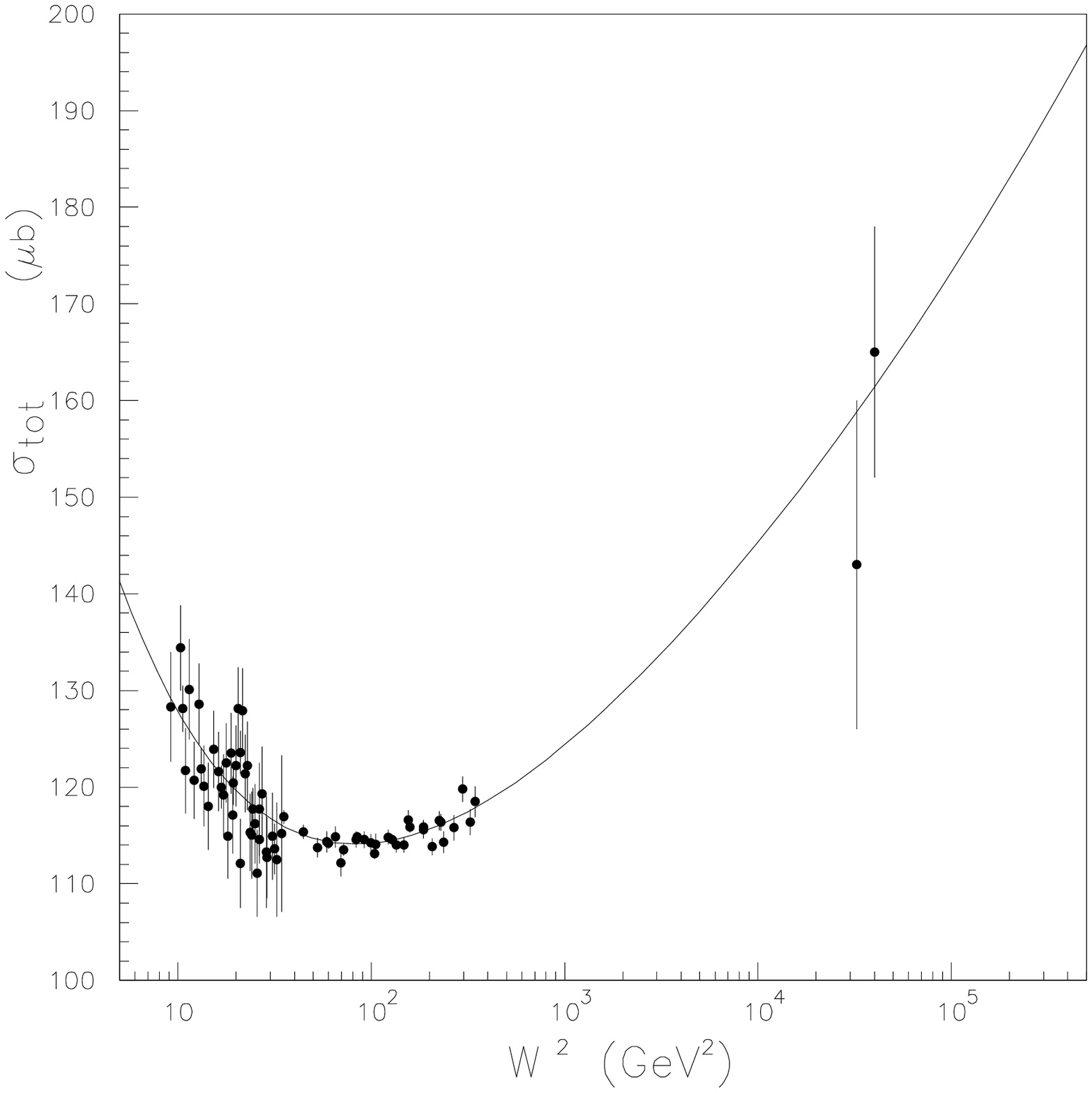}

Fig. 4
\end{center}

\begin{center}
\leavevmode\epsfbox{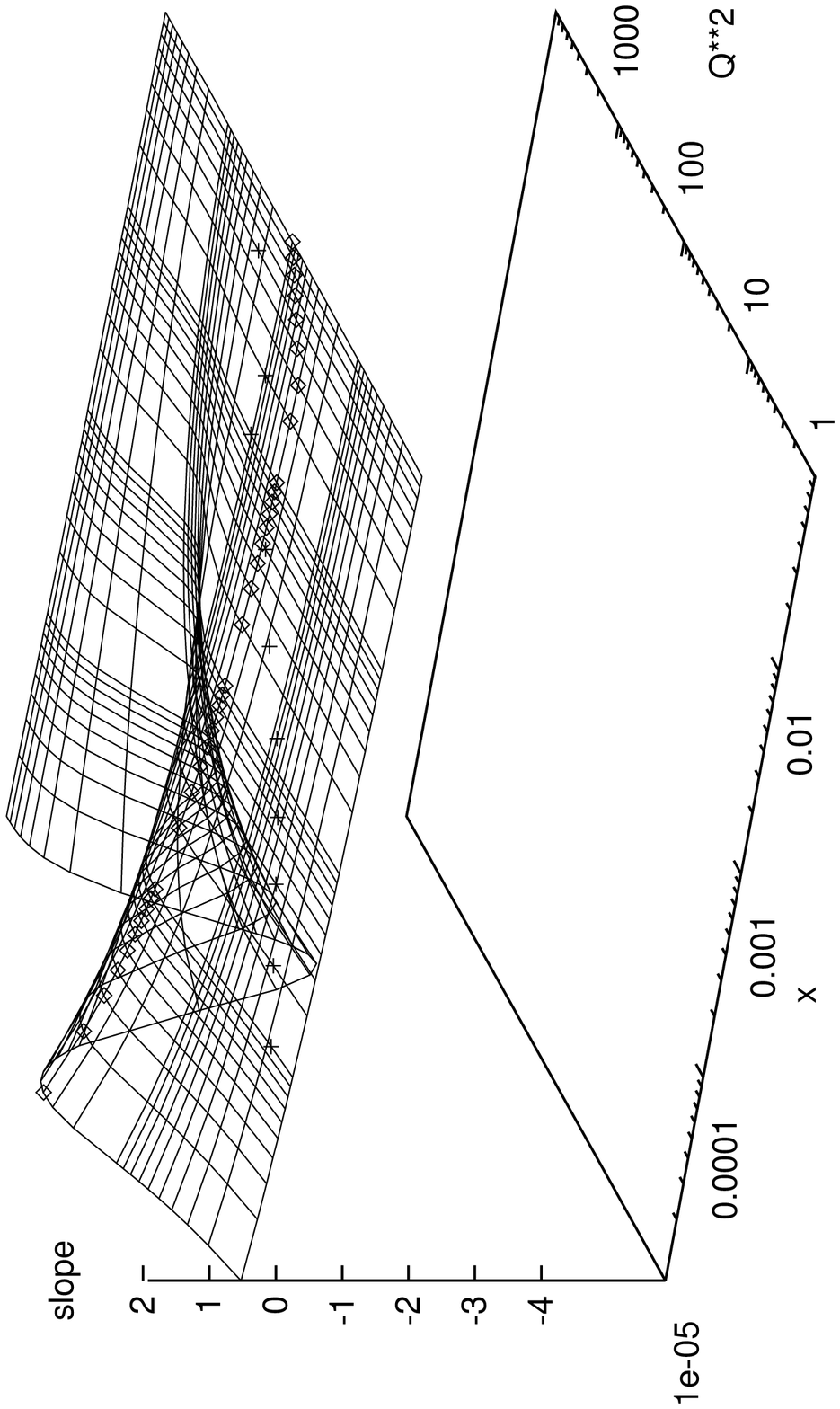}

Fig. 5
\end{center}

\begin{center}
\leavevmode\epsfbox{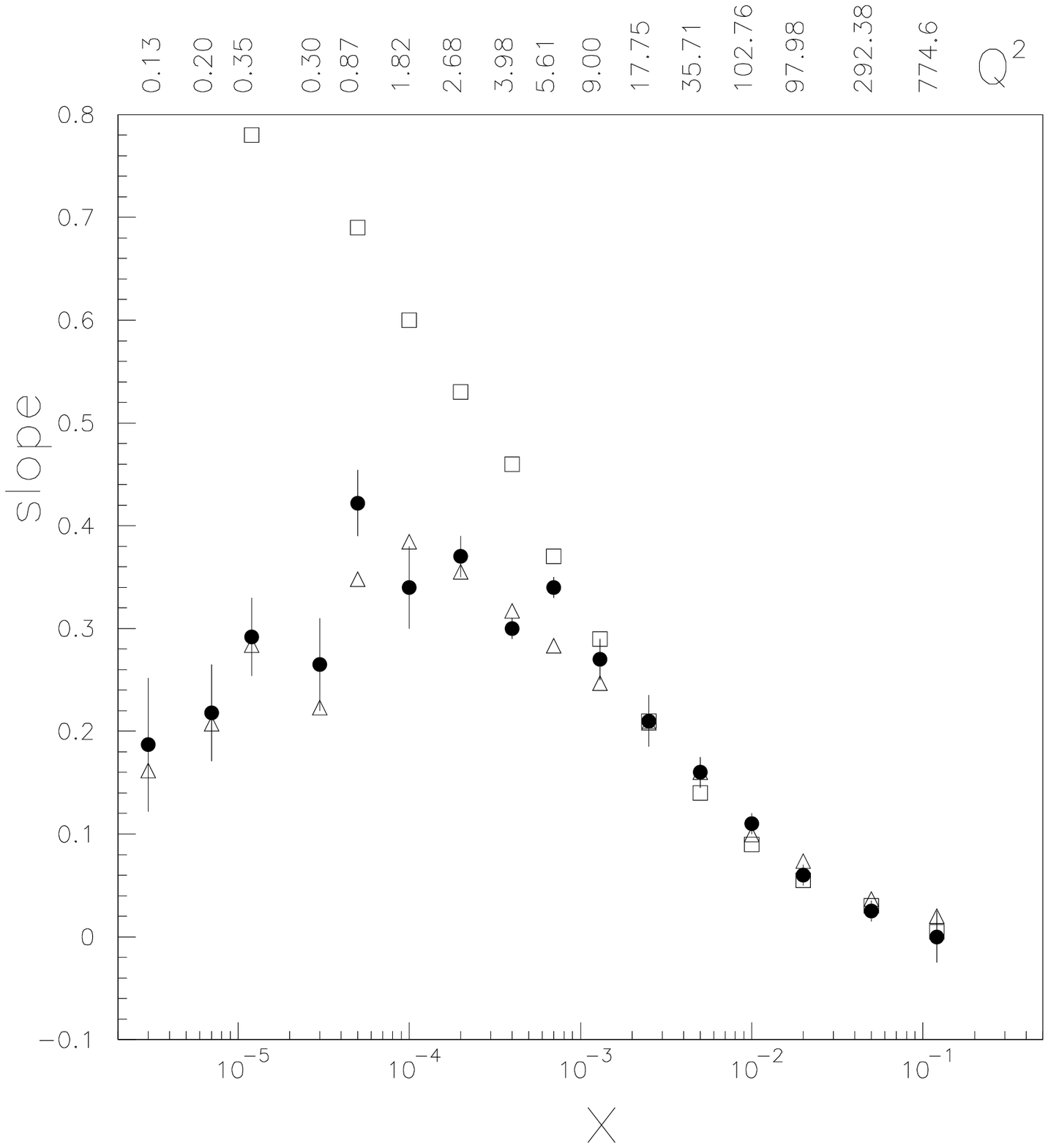}

Fig. 6a
\end{center}

\begin{center}
\leavevmode\epsfbox{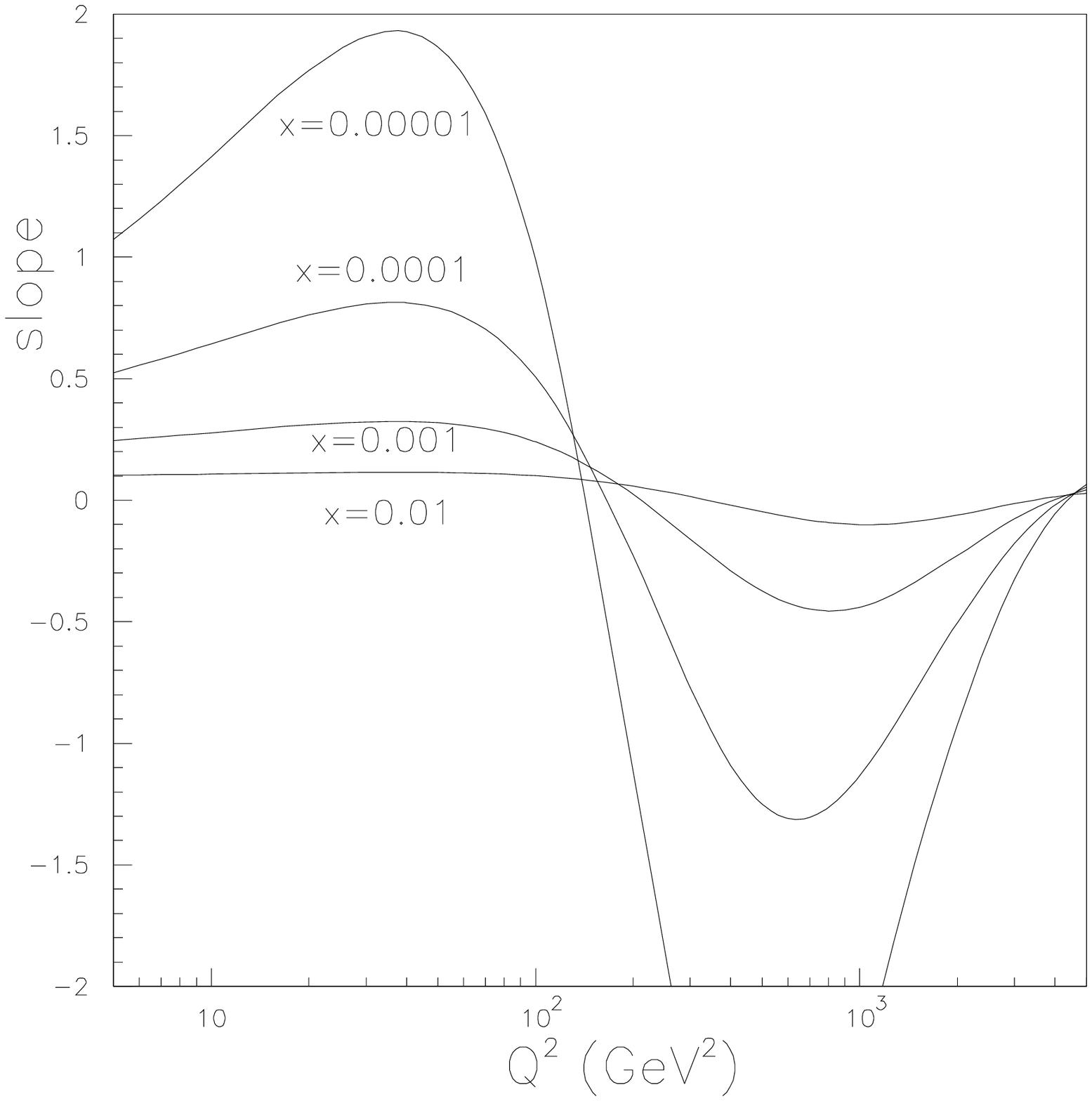}

Fig. 6b
\end{center}

\begin{center}
\leavevmode\epsfbox{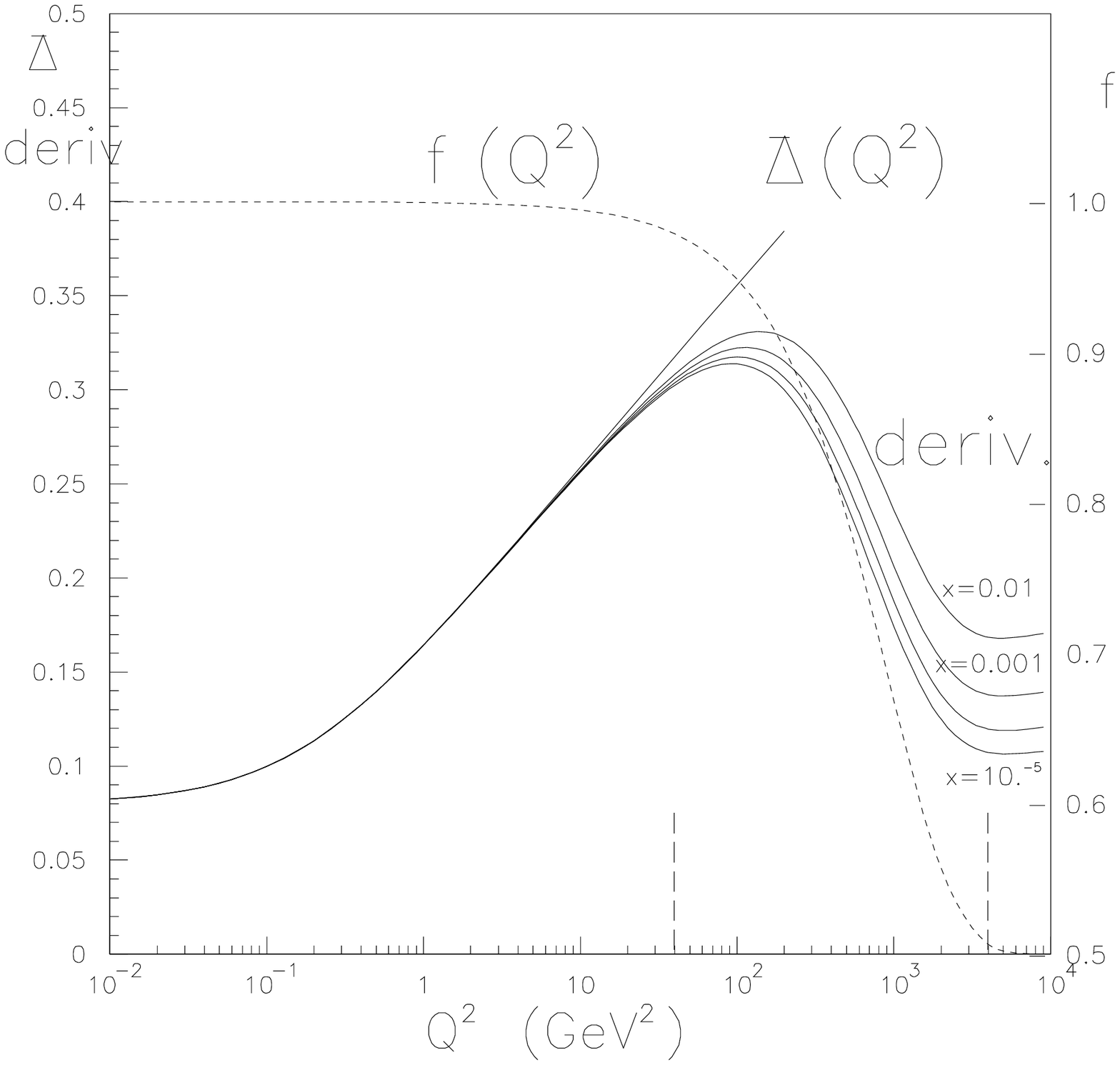}

Fig. 7
\end{center}
\end{document}